# A Passive UV-Based System for Discharging the Test Masses of Ground-Based Gravitational Wave Detectors

S. Buchman[1], S. Wang[2], S. Saraf[3]

[1]Hansen Experimental Physics Laboratory, Stanford, CA 94305 USA
[2]Hainan Tropical Ocean University, Sanya, 572022 China
[3]SN&N Electronics, Inc., 1846 Stone Avenue, San Jose, CA 95125 USA

**ABSTRACT**

We present a passive ultraviolet (UV) charge management system for the fused silica test masses (TMs) in ground-based laser interferometric gravitational wave detectors. The system uses photoelectron emission from low work-function gold (Au) coatings illuminated by 275–285 nm UV light to neutralize unwanted electric charge on electrically floating TMs, maintaining them at zero potential relative to surrounding components. Two implementation schemes are described: (1) distributed discharge tabs illuminated by UV LEDs mounted at the eight surrounding earthquake stops, and (2) a conductively linked tab design enabling centralized discharge on the TM barrel, with potential extension to annular coatings around the high-reflectivity and anti-reflective surfaces. Experimental results show photoelectric currents $\geq$10 pA for 1.0 mW of incident UV, enabling discharge rates $\geq$10 V/s for a 1 pF TM. With 0.2 mW UV power, charge neutralization to $\leq$1 pC can be achieved in 5–75 minutes. This technique offers a vacuum-compatible alternative to ion sprayers, significantly reducing operational downtime. For the conductive-barrel configuration, we also propose a balanced electrostatic actuator using four parallel-plate capacitors for combined control of TM axial displacement, tilt, and azimuth, with reduced drive voltages.

# I. OVERVIEW

Gravitational waves (GWs) were first detected[1] and to date remain only observable using ground-based laser interferometric detectors, such as the LIGO[2], VIRGO[3], GEO600[4], and KAGRA[5] observatories, whose simplified schematics is illustrated in Figure 2. The fused silica ($SiO_2$) end mirrors, commonly referred to as test masses (TMs), form the resonant optical cavities of the interferometer and are isolated from external perturbations, including seismic motions, thermal fluctuations, residual gas motion, and electrostatic forces. Figure 2 illustrates the multi-stage suspension chain used to attenuate seismic disturbances and mechanically isolate the TM from ground motion and the reaction mass (RM). The TM has a high reflectivity (HR) coating facing the optical cavity and an antireflecting (AR) coating on the RM side.

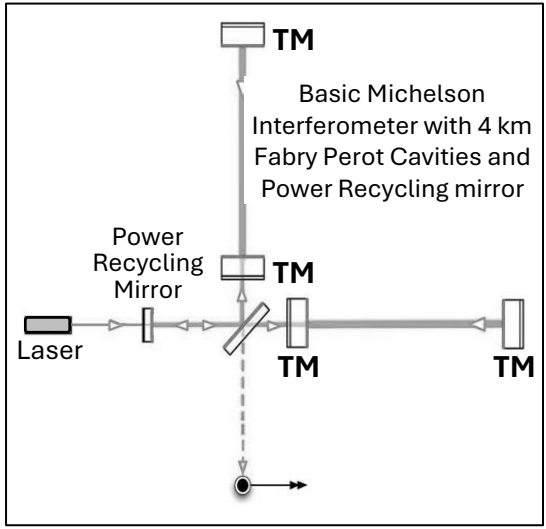

Figure 2. Schematic of GW detector

Electrostatic charging of the TM in the GW detectors occurs primarily via the triboelectric effect[6] by contact with the earthquake stops (ESs)[7] shown in Figure 3. Other significant charging mechanisms are the friction with the flowing gas during system pump down, electrons generated by the ion pumps, and the unbalanced electrostatic drive (ESD). Smaller charging effects are due to cosmic radiation[8] and ionized residual gas[9].

Charges on the TM reduce its $Q$ factor[10] with additional effects due to their motion across the HR surface and their interaction with the ESD field[11] and with the movement of the surrounding support structures[12]. For Advanced LIGO the allocated total fluctuating TM charge, consistent with its noise budget, is $\cong 10pC$ or $\cong 16fC/cm^2$ surface density, and has a decay time of about 1 year[9] (TM area of about 500 $cm^2$). Lanz[12] estimates that the motion of the charged TM, caused by the moving of the grounded surrounding metal structure, exceeds the error budget for a surface density of $\sim 200pC/cm^2$ or about $10^4$ higher than for the fluctuating charge mechanism. Therefore, the requirement for the TM charge is $\leq 10pC$, with a goal of $\leq 1pC$.

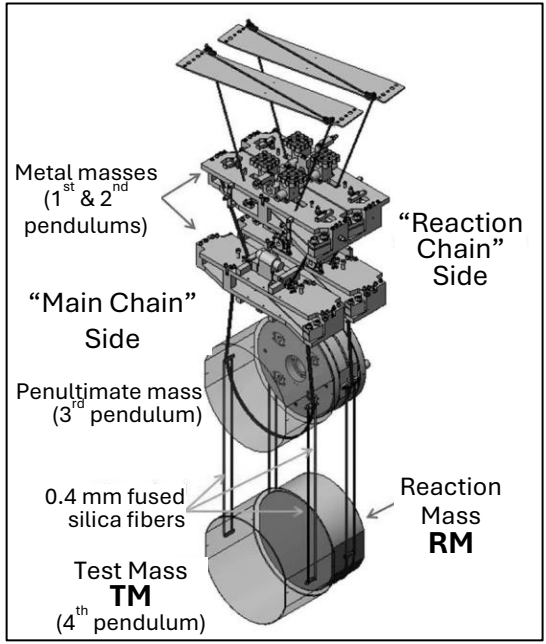

Figure 2. TM multistage suspension chain

Presently, the charge mitigation system for LIGO is based on two techniques[13]. The first system is passive and uses $SiO_2$ ESs that contact the bare $SiO_2$ areas of the TMs during anomalous large motions. While small scale laboratory tests indicate that no charge is transferred during $SiO_2$-to-$SiO_2$ contact[14], in practice actual earthquakes in LIGO do result in significant charging of the TMs[15], thus invalidating this approach. In order to offset this charge, LIGO successfully uses a secondary active system based on the concept of a low-pressure ion sprayer[16,17] implemented by

R. Weiss[18,19]. Low pressure nitrogen gas is ionized in an electric field, thus producing both electrons and positive ions[20]. The charged TM preferentially attracts charges of opposite polarity, while like charges are neutralized on the surrounding grounded surfaces. Operationally, however, this system has the significant disadvantage of interrupting science data acquisition for several days; it requires isolating the detectors' vacuum chambers, activating the ionized gas discharge, re-establishing the system vacuum, and re-initializing the interferometric lock. Note that charge management using photoelectrons was demonstrated for GEO600, albeit in a different configuration and control mode[21].

Here, we propose replacing or supplementing the two currently employed LIGO charge management techniques with an alternative, non-contact method using UV-generated photoelectrons[22] - a technology validated by the Gravity Probe B[23], UV-SaudiSat[24,25], and LISA Pathfinder[26] missions. In particular, we propose the technique pioneered by Wang *et. al.*[27] of using UV wavelengths in the range of 275-285 nm to generate 'slow photoelectrons' from Au films enabling passive charge management without the need for bias voltages. The energy of the UV (4.35 eV-4.51 eV) is sufficiently close to the maximum of the Au work function distribution[27,28], ~4.3 eV causing the photoelectron flow to be dominated by the potential difference between the two film surfaces, thus creating a virtual short between them. This method minimizes disturbances to the test masses (TMs) and can be continuously activated.

Two design implementations are presented. The first integrates UV LEDs at each of the eight ESs surrounding the TM, illuminating small-area, low-work-function coating tabs positioned on the TM surfaces. A second approach extends this design to mitigate dielectric surface charging by applying a low-conductivity coating to the TM barrels, connected to the ES tabs, with UV LEDs illuminating the barrel directly. In both implementations the ESs are conductive and grounded. Note that the UV irradiation causes the deterioration of the HR coating and therefore needs to be largely limited to the photoemission area[29]. The main goals we address are:

1) A non-contact charge management approach, based on UV emitted photoelectrons, that requires no system vacuum breaking.
2) Minimizing electrostatic charging and forces acting on the TMs, RMs, and ESs by covering their dielectric surfaces with conductive coatings.
3) Preventing imbalances or unwanted electrostatic forces on the TMs and the RMs by balancing the electrostatic drive (ESD) so as the potential induced on them is zero:

$$\sum_{i=1}^{n} C_i [V_{DC} + V_{AC}(t)]_i = 0 \qquad 1$$

Where $n$ is the number of different conducting surfaces of the ESD and $C_i$, $(V_{DC})_i$, and $[V_{AC}(t)]_i$ their respective capacitances to the TM and the DC and AC voltages applied by the ESD.

The layout of the work is as follows. Section II describes the system based on directing the UV LED radiation into the ESs, by either integrating the diodes directly into the ESs or channeling their power through optical fibers. Section III details the approach involving using a common set of UV-LEDs, while the TM tabs are conductively connected by either thin film strips or by coating the entire barrels of the TMs. Here we also discuss the advantages of conductive coatings on the dielectric surfaces of the TM, the barrel and the ESD-facing annulus. Section IV contains the conclusions of the paper, while a proposed implementation for the required changes to the ESD is shown in Appendix A.

## II. SYSTEM WITH UV POWER DIRECTED INTO EACH EARTHQUAKE STOP

In this implementation, we replace the existing passive LIGO configuration of $SiO_2$ ESs facing bare $SiO_2$ surfaces of the TMs, with conductive and grounded ESs integrated with coaxial UV-LEDs directed at conductive discharge tabs coated on the TMs. This active charge management approach enables the controlled discharging of the TMs following contact with any ES, such as after a seismic event, thereby mitigating residual electrostatic forces without requiring the time-consuming nitrogen ion spray procedure as for the currently used, active, LIGO discharge system. Figure 3 is a CAD representation of the TM, the RM, the ESs, and their support structure with the four forward ESs on the opposite side of the RM facing the optical cavity of the interferometer.

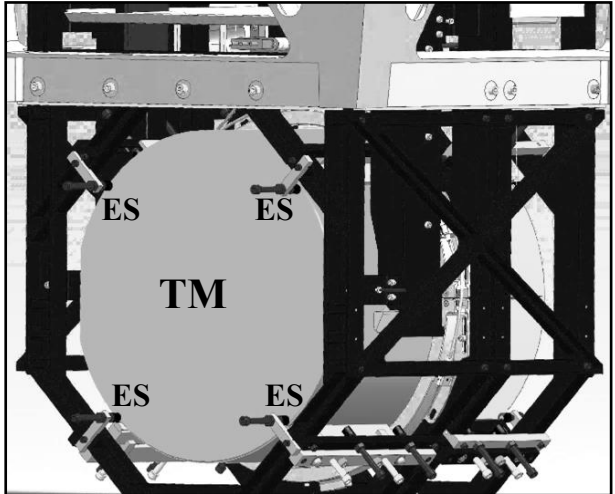

Figure 3. CAD representation of TM (gray), RM (light gray), ESs (labeled), and support structure (black).

The ESs of LIGO are mounted on screws of approximately 12 mm diameter. Figure 4 (not to scale) shows a schematic cross section for two practical approaches to directing the UV light into the ESs; a) has the LED mounted directly in the stop, while in b) the UV light is routed through the stop by a UV compatible fiber. Gray solid areas represent the ES metal, with the $SiO_2$ of the TM shown patterned, and the films for photoemission and binding to $SiO_2$ in black and gray respectively. The ESs are grounded and, in order to prevent damage caused by like-metal contact, their sections that can touch the TMs are left uncoated and facing the $SiO_2$ binding film. An elastic conductive component, that is represented by spring shapes, supports the forward section of the ESs and is required to reduce impact forces during earthquakes.

In this implementation ES to TM contact is between grounded ES metal to $SiO_2$ binding film, replacing the present design of contact between insulated $SiO_2$ ES tips to TM $SiO_2$, while a conducting elastic support (spring, Belleville washer stack) replaces the presently used insulating rubber. Note that the Au film surfaces are confined to the photoelectric discharge areas.

Laboratory measurements and ray tracing simulations will direct the choice of shape for the TM facing areas of the ESs, such that the UV flux is largely confined during multiple reflections. We expect that conical, parabolic, or parallel profiles will all be acceptable. Criteria for the choice of film coating materials for the TM tabs and the ESs are the following: low photoelectric work function, good coating adhesion to the TM $SiO_2$ surface, moderate to high compressional strength, low magnetic susceptibility, and, most importantly, previous performance in similar applications. Table 1 gives a list of metal film choices, with their key properties and performance regarding these criteria, as candidate materials for UV photoemission films for LIGO. Note that the value for the Au work function is taken from relevant devices and experiments[27,28]. Mo is the least promising material due to its somewhat higher work function, poor adhesion to $SiO_2$ and lack of heritage. Cr and Ti meet all requirements but will require laboratory validation; with no relevant data for Cr and disappointing data for Ti in the GP-B experiment[22] where it had to be coated with Au for enhanced and consistent photoemission. As coating for the rotors of the GP-B gyroscopes Nb photoemitted very well under 254 nm illumination[22] and should be rated high as a choice of material for the UV charge management films for LIGO.

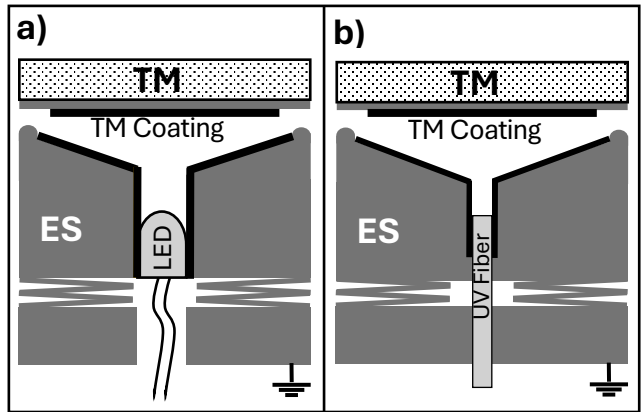

Figure 4. Two approaches for directing the UV light into the ESs. a) UV-LED in ES, b) UV fiber in ES. Au coating (black), Ti/Cr coating (dark gray)

Au is the material with the greatest amount of experimental data from the GP-B[22,23], UV-SaudiSat[24,25], and LPF[26] flight missions and from extensive laboratory work[27,28]. In particular, the proposed passive charge management approach was demonstrated on Au coated surfaces[27,32]. The poor bonding to $SiO_2$ and low hardness of Au can be overcome by coating a larger area of a bonding film for the Au tabs on the TMs – for example Cr or Ti– and applying an Au film on the UV reflector area of the ESs. Contact during earthquakes will thus occur between the ES metal and the tab adhesion layer film on the TM, preventing damage to the Au films.

Table 1. Essential attributes of candidate materials for UV photoemission films

| Material | Work function (eV) | Adhesion to $SiO_2$ | Hardness (Mohs scale) | Magnetic susceptibility (dimensionless) | Previous UV performance |
|---|---|---|---|---|---|
| Cr | 4.50 | Excellent | 8.5 | $+317.7\times10^{-6}$ | N/A |
| Mo | 4.65 | Poor | 5.5 | $+120.3\times10^{-6}$ | N/A |
| Nb | 4.41 | Excellent | 6.0 | $+237.0\times10^{-6}$ | Proven |
| Ti | 4.33 | Excellent | 6.0 | $+180.7\times10^{-6}$ | Not proven |
| Au | 4.1-4.6 | Poor | 2.5 | $-34.4\times10^{-6}$ | Proven |

The appropriate range of UV-LED wavelengths for passive charge management using 'slow' photoelectrons emitted from Au films, is 275-285 nm (4.51-4.35 eV)[27,32]; a range covered by a large number of UV-LEDs from various vendors. Table 2 shows a few of the suitable devices, their manufacturers, wavelength, power outputs, and maximum diameters.

Table 2. A few examples of LED-UVs in the wavelength range of 275-285 nm.

| LED Mounting | Manufacturer | Wavelength (nm) | Power at 10mA (mW) | Diameter (mm) | Remarks |
|---|---|---|---|---|---|
| Micro-LED Lens | Development[30] | 274, 282 | 0.023, 0.017 | 5.6 | Square mount |
| Micro-LED Flat | Seoul Viosys[31] | 275 | 2.0 | 5.0 | Square mount |
| Micro-LED Array | Development[32] | 275 | 0.5 | 0.05-0.30 | Mounting question |
| TO-39 Ball Lens | Marktech Optoelelectronics[33] | 275-285 | 1.0 | 9.1 | 60mW power dissipation |
| TO-46 Ball Lens | | 280 | 1.0 | 5.45 | |
| Bolb S3535 | LedRise[34] | 275 | 40 (200mA) | 5.0 | 1.5W power dissipation |

For the UV-LEDs mounted into the ESs, as in Figure 4a, the critical properties are maximum diameter, convenience of mechanical integration, and power dissipation. Larger LED cases reduce the available reflective area of the present 12 mm diameter ESs; by a significant 56% for the TO-39 and an acceptable 22% for the TO-46 device respectively. Use of micro-LEDs will require specialized manufacturing with mounting and connections adapted to this application. The maximum TM temperature increase $\Delta T$, assuming it continually absorbs the entire UV power, $P_{UV}^{MO} = 1\text{mW}$, of the TO-46 Marktech Optoelectronics LED, is given by:

$$\Delta T < \frac{8 P_{UV}^{MO}}{4\sigma_{SB} T^3 A_{TM} \epsilon_{TM}} \cong 6\text{mK} \qquad 2$$

where $\sigma_{SB} = 5.67 \times 10^{-8}\ \text{W/m}^2\text{K}^4$ is Stefan-Boltzmann's constant, and the TM's temperature, surface area, and emissivity are $T = 300\text{K}$, $A_{TM} = 0.3\text{m}^2$, $\epsilon_{TM} \approx 0.7$. We note that a maximum TM temperature increase of $\leq 6\text{mK}$ is well below the allowed variation. The much larger thermal power generated by the LED's, $P_{LED} \cong 0.5\text{W}$, will be dissipated by conduction through the metal support structure, rather than by radiation for the TM, and thus will not constitute a problem.

Depending on the lenses (or flats) mounted in front of the LEDs, the radiation angles, $2\theta_{1/2}$, vary between 12 and 120 degrees. Due however to the large aspect ratio of the ES diameter to its distance to the TM, approximately 12 mm to 2 mm, the radiation angle should not constitute a driving consideration in the choosing of the UV source. The UV-LEDs can be fabricated to include photodetectors in standard TO type cans or flat mounts and fronted by lenses or flat windows[25]. The photodetector is a duplicate LED, therefore serving both as a diagnostic tool and as a backup for the primary device. These types of devices were used in the UV-SaudiSat flight mission[24,25].

$dV_{TM}/dt$, the time dependent decay of the TM potential, caused by the photoelectron current $I_{PE}$, is characterized by two different regimes:

$$|V_{TM}| > q^{-1}(hc/\lambda_{LED} - \phi_{Au}) \equiv \mathbb{V}_{PE} \qquad 3a$$

$$|V_{TM}| < q^{-1}(hc/\lambda_{LED} - \phi_{Au}) \equiv \mathbb{V}_{PE} \qquad 3b$$

where $q, h, c, \lambda_{LED}$, and $\phi_{Au}$ are the electron charge, Planck's constant, the speed of light, the UV-LED wavelength, and the Au work function, while the ESs are grounded. We defined the transition potential between the modes as $\mathbb{V}_{PE}$, where $\mathbb{V}_{PE} > 0$ to enable photoemission. As photoelectrons are generated from both the TM and the ES, the two cases in equation 3 each divide into two values, though of similar functional form. For $|V_{TM}| > \mathbb{V}_{PE}$:

$$0 < \mathbb{V}_{PE} < V_{TM} \Rightarrow \frac{dV_{TM}^R}{dt} = -\frac{1}{C_{TM}}\frac{dQ_{TM}}{dt} = -\frac{I^R}{C_{TM}} \qquad 4a$$

$$0 > -\mathbb{V}_{PE} > V_{TM} \Rightarrow \frac{dV_{TM}^D}{dt} = \frac{1}{C_{TM}}\frac{dQ_{TM}}{dt} = \frac{I^D}{C_{TM}} \qquad 4b$$

$Q_{TM}$ and $C_{TM}$ are the TM charge and its capacitance to be nullified. In equation 4a the photocurrent from the TM is blocked $I^D = 0$, while $I^R$ is the photoelectric current generated from the ES under reflected UV illumination (superscript $R$), while equation 4b illustrates the opposite case: $I^R = 0$ and the photoelectron current is generated from the direct illumination of the TM (superscript $D$). In the $|V_{TM}| > \mathbb{V}_{PE}$ regime $dV_{TM}/dt$ is linear and proportional with $I^D$ or $I^R$, with $I^D > I^R$ in general and $I^D/I^R \cong 20$ for photoemission from Au parallel plates at 275 nm[27,32].

In the $|V_{TM}| < \mathbb{V}_{PE}$ case the photocurrents $I_{SS}$ and $I_{TM}$ flow simultaneously and in opposite directions across the gap from the ES to the TM, with the net current $I_{gap}$ given by:

$$I_{gap} \equiv I_{SS} - I_{TM} = V_{TM} \mathcal{B}_{LED} \mathcal{F}(\mathfrak{R}_{SS}^R, \mathfrak{R}_{TM}^D) = \frac{V_{TM}}{\mathbb{R}_{PE}^{D,R}} \qquad \mathbb{R}_{PE} \equiv [\mathcal{B}_{LED}\mathcal{F}(\mathfrak{R}_{SS}^R, \mathfrak{R}_{TM}^D)]^{-1} \qquad 5$$

$I_{gap}$ is proportional to $V_{TM}$ and inverse proportional to the equivalent resistance $\mathbb{R}_{PE}$ of the photoelectron 'wire' that is a function of $\mathcal{B}_{LED}$, the performance and functional parameters of the UV-LED, and on $\mathfrak{R}_{ES}^{R}$ and $\mathfrak{R}_{TM}^{D}$, the optical properties and quantum efficiencies of the photoelectric emitting films on the ES and its facing tab on the TM; see Figure 4.

For $|V_{TM}| < \mathbb{V}_{PE}$:

$$0 < V_{TM} < \mathbb{V}_{PE} \Rightarrow \frac{dV_{TM}}{dt} = -\frac{1}{C_{TM}} \frac{dQ_{TM}}{dt} = -\frac{I^R}{C_{TM}} = -\frac{V_{TM}}{C_{TM}\mathbb{R}_{PE}^R} \qquad 6a$$

$$\Rightarrow V_{TM}(t) = V_{TM}^0 e^{-t/\tau^R} \qquad \tau^R \equiv C_{TM}\mathbb{R}_{PE}^R$$

$$0 > V_{TM} > -\mathbb{V}_{PE} \Rightarrow \frac{dV_{TM}}{dt} = \frac{1}{C_{TM}} \frac{dQ_{TM}}{dt} = -\frac{I^D}{C_{TM}} = -\frac{V_{TM}}{C_{TM}\mathbb{R}_{PE}^D} \qquad 6b$$

$$\Rightarrow V_{TM}(t) = V_{TM}^0 e^{-t/\tau^D} \qquad \tau^D \equiv C_{TM}\mathbb{R}_{PE}^D$$

From equation 6 we obtain the expected time dependence, equivalent to a resistor-capacitor (RC) circuit, with time constant $\tau^D$ or $\tau^R$:

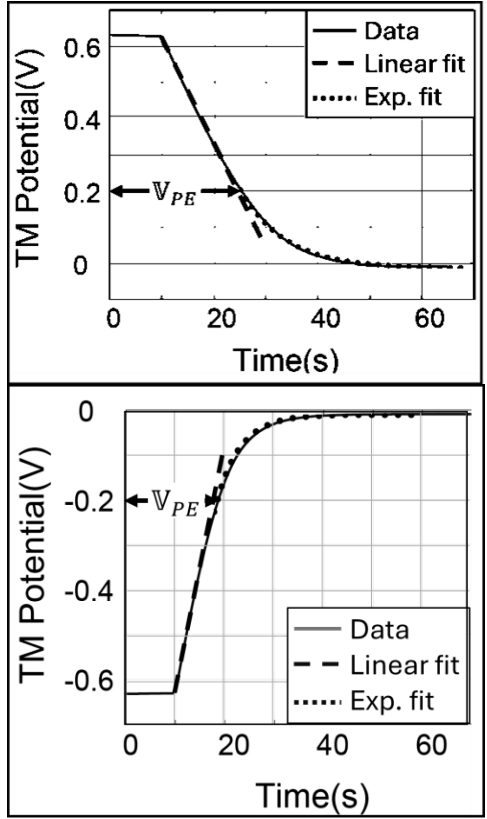

Figure 5. Discharge data using 275 nm UV on Au coatings: top $V_{TM} > 0$ and bottom $V_{TM} < 0$, showing linear and exponential fits for $V_{TM} \gtrless \mathbb{V}_{PE}$ respectively

$$\frac{dV_{TM}(t)}{dt} = -\frac{V_{TM}(t)}{\tau} \qquad 7$$

here $V_{TM}^0 \equiv V_{TM}(t=0)$. Due to the asymmetry between the direct and reflected illumination constants ($R\ and\ D$ superscripts) the time constants $\tau^R$ and $\tau^D$ in equations 6a and 6b are in general not equal. While the data from references 27 and 32 validate the model of the TM potential decay as linear followed by an exponential, we note that the transition between the regimes is gradual as both $\lambda_{LED}\ and\ \phi_{Au}$ have gaussian distributions around their peak values[27]. Figure 5 shows that the discharge data from Au films using a 275 nm UV-LED[27,30] is linear for $|V_{TM}| > 0.2V$ and exponential for $|V_{TM}| < 0.2V$.

We approximate $C_{TM}^{SS}$, the capacitance between the ES and the tab on the TM, as a parallel plate disk capacitor of 6 mm radius and 2 mm gap: $C_{TM}^{SS} \cong 0.5pF$. As $\mathbb{V}_{PE} \approx 0.2V$ for $\lambda_{LED} = 275nm$ and photoemitter Au films, the main part of the TM discharge occurs linearly, with $I_{SS}^R \cong 0.5pA$ (reflected UV illumination) and $I_{TM}^D \cong 10pA$ (direct UV illumination); for an LED producing $P_{UV} \cong 1.0mW$. For $|V_{TM}| < \mathbb{V}_{PE}$, the discharge time constants are: $\tau^R \leq 1.5s$ and $\tau^D \leq 0.2s$. In Table 3 columns 2 to 5 - using equations 4 and 6 and the values above[27,32] - we summarize the discharge rates for one ES LED, in the four $V_{TM}$ charge ranges; for $C_{TM}^{SS} \cong 0.5pF$ and $\mathbb{V}_{PE} = 0.2V$.

The discharge durations from a TM charge of $\pm\ 0.5$ nC ($\pm\ 1$ kV potential estimated as a plausible high value) to $|V_{TM}| = \mathbb{V}_{PE} = 0.2V$ are in the linear regime 1,000 s and 50 s, while the $3\tau$ times required to further reduce the potential by 95% to 10 mV are 2.3 s and 0.3 s. After discharging to $|V_{TM}| \leq 10$mV at each of the 8 ESs, the total TM charge $|Q_{TM}| \leq 0.04$pC $\ll 1$pC, a factor of 25 lower than the goal of 1 pC and a factor of 250 lower than the Advanced LIGO error budget[9].

**Table 3. Discharge rates and time constants for the UV-LEDs mounted in the ESs or in single discharge.**

| $V_{TM}$ (V) | ES to TM discharge $P_{UV}^{MO} = 1$mW, $C_{TM}^{SS} \cong 0.5$pF | | | Single discharge to TM tabs $P_{UV}^{LR} = 40$mW, $C_{TM}^{\langle\varepsilon\rangle} \cong 20$pF | | | Single discharge to coated TM $P_{UV}^{LR} = 40$mW, $C_{TM}^{T} \cong 100$pF | | |
|---|---|---|---|---|---|---|---|---|---|
| | $dV_{TM}/dt$ (V/s) | $dQ_{TM}/dt$ (pC/s) | $\tau$ (s) | $dV_{TM}/dt$ (V/s) | $dQ_{TM}/dt$ (pC/s) | $\tau$ (s) | $dV_{TM}/dt$ (V/s) | $dQ_{TM}/dt$ (pC/s) | $\tau$ (s) |
| $V_{TM} > 0.2$ | -1.0 | -0.5 | N/A | -1.0 | -20.0 | N/A | -0.2 | -20.0 | N/A |
| $0 < V_{TM} < 0.2$ | N/A | N/A | 0.75 | N/A | N/A | 0.75 | N/A | N/A | 3.75 |
| $-0.2 < V_{TM} < 0$ | N/A | N/A | 0.1 | N/A | N/A | 0.1 | N/A | N/A | 0.5 |
| $V_{TM} < -0.2$ | 20.0 | 10.0 | N/A | 20.0 | 400 | N/A | 4.0 | 400 | N/A |

Note however, that discharging the ESs' facing tabs does not address the issue of accumulation of electric charges on the dielectric surfaces of the TM. Discharge times to $|V_{TM}| \leq 10$mV are about 15 min and 1 min for $V_{TM} > 0$V and $V_{TM} < 0$V respectively. As these are upper limit estimates, it follows that about 0.2 mW of UV optical power output at the ESs should be suitable for this application.

While embedding the LEDs in the ESs maximizes the UV power into the discharge space, it has the disadvantage of placing active devices in close vicinity to the TMs. Alternately, the UV power can be delivered to the ESs via optical fibers, Figure 4b, by using UV-optimized quartz fibers that conduct 275 nm wavelength with no darkening and only about 0.25 dB/m power loss[35]. Using multi core fiber cables (e.g. 7 fibers of 200 μm each for an approximately 2 mm diameter) provides redundancy against breakage and allows for a tighter bend radius (of about 5 cm) than that of a single larger diameter fiber[36]. The UV-LEDs will then be located outside the vacuum chamber of the TM, thus reducing complexity, allowing easy access to all the active components of the system, and removing possible noise sources from the vicinity of the high sensitivity elements of the GW detector. We suggest three options for the coupling of the UV light into the fiber optics cables:

1) Aligning the fiber to an LED of about 1 mW with a lens window[25] for maximum coupling. The total loss from coupling to the LED and a 3 m length of fiber is less than 5 dB, resulting in about 0.3 mW optical power at the ES.

2) Using a high-power LED, (e.g. Bolb S3535 UVC SMD LED 275 nm 40 mW)[34] coupled to all of the 8 fiber optic cables. With 40 mW of optical power at 200 mA operational current and a wide radiation angle this LED would provide more than 0.4 mW at each of the ESs and adequately meet the TM discharge requirements in less than 1 hour.

3) A custom designed LED coupled to each of the fiber cables that would have the advantage of ease of installation and calibration, with the drawback of difficulty of replacement of the LED. Inserting an additional fiber optics connector between the LED coupled fiber and the fiber cable to the ES would overcome this disadvantage, while reducing the useful optical power by only an acceptable ≤ 3 dB.

## III. SYSTEM WITH CONNECTED TM TABS AND COMMON UV SOURCE

Conductively connecting the TM tabs allows for a single UV source located on the support structure. Figure 6 shows a conceptual cross section of the ESs and tabs for this implementation. The TM tabs are coated with only a binding film of Ti or Cr, while the ESs remain uncoated to ensure non-similar metals contact during earthquakes, thus avoiding adhesive wear.

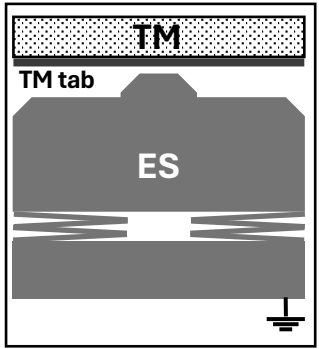

**Figure 6. Cross section of ES and TM for connected TM tabs configuration**

Depending on the ease of implementation, two versions of the thin trace conductive film connection of the TM tabs are suggested and schematically illustrated in Figure 7, with the binding Ti or Cr layer shown in gray and the Au photoemission film in black:

1) Four linear traces linking opposite tabs across the TM barrel, and connected together by a circular median of the barrel: Figure 7a.
2) Two rings joining each of the four tabs on the TM edges, facing the RM and the beam sides respectively, that are then connected along the bottom of the TM barrel: Figure 7a.

For both configurations we propose a parallel Au coated 3 mm discharge gap of $\cong 3$ cm² area, with a capacitance $C_{Au} \cong 1$ pF.

A UV-LED or a single optical fiber cable can supply power to the discharge, similarly to the implementation with sources in each ES. Redundancy will be achieved, again likewise to the ES imbedded sources, by incorporating a photodetector identical to the main LED or by a multiple-fiber optical cable. In the following, we show that as the single UV source a conservative choice is a higher power device, e.g. Bolb S353534 with $P_{UV}^{LR} = 40$mW optical power output[34].

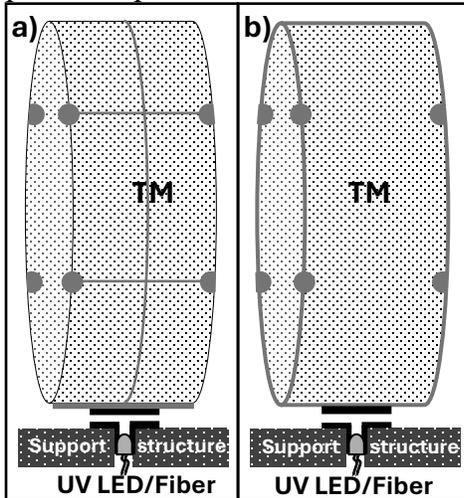

**Figure 7. Two thin trace versions for the TM tabs barrel connection (Ti/Cr dark gray, Au black): a) one ring and four across b) two rings and one across.**

In these configurations the capacitance to be discharged, $C_{TM}^0$, is between the mostly uncoated TM (34 cm diameter, 20 cm thickness) and its surrounding structures, that we will approximate as a circular single plate capacitor, $C_{SP}$, to the RM with the dielectric constant of SiO₂, $\varepsilon_Q = 3.6$, and the radius, $r_{RM} = 17$cm. As $\langle \varepsilon \rangle < \varepsilon_Q$ and the ESD conducting area is much smaller than the total RM area, we estimate $C_{TM}^{\langle \varepsilon \rangle}$, the capacitance of the TM to ground, to be about $C_{SP}/3$:

$$C_{SP} = 8\varepsilon_Q \varepsilon_0 r_{RM} = 45\text{pF} > C_{TM}^{\langle \varepsilon \rangle} \cong 20\text{pF} \qquad 8$$

$dV/dt$ and $1/\tau$ scale proportionally to $P$, the optical power of the UV-LEDs, and inverse proportionally to $C$, the capacitances to be discharged, while $dQ/dt$ is proportional only to the optical power:

$$\frac{(dV/dt)_2}{(dV/dt)_1} = \frac{(1/\tau)_2}{(1/\tau)_1} = \frac{P_2}{P_1}\frac{C_1}{C_2} \,, \quad \frac{(dQ/dt)_2}{(dQ/dt)_1} = \frac{P_2}{P_1} \qquad 9$$

Columns 5 to 7 of Table 3 show the values for $dV/dt$, $dQ/dt$, and $1/\tau$ for $P_{UV}^{LR} = 40$mW and $C = 20$pF. Note that the coincidental equality $P_{LED}^{LR}/P_{LED}^{MO} = C_{TM}^{\langle \varepsilon \rangle}/C_{TM}^{ES}$ causes thee values of $dV/dt$ and $\tau$ to be equal to those for the ES to TM discharge. The durations for discharging the TM to $|V_{TM}| \leq 10\text{mV}$ are approximately 60 s and 4.0 s for $Q_{TM}^0 = \pm 1\text{nC}$ respectively ($V_{TM} = 50\text{V}$). $\Delta T < 30\text{mK}$ is the maximum temperature increase of the TM as calculated from equation 2. Similar to the ES-imbedded discharge implementation, the UV power can be directed to the single discharge via a fiber optic cable with a power loss of $\cong 5\text{dB}$.

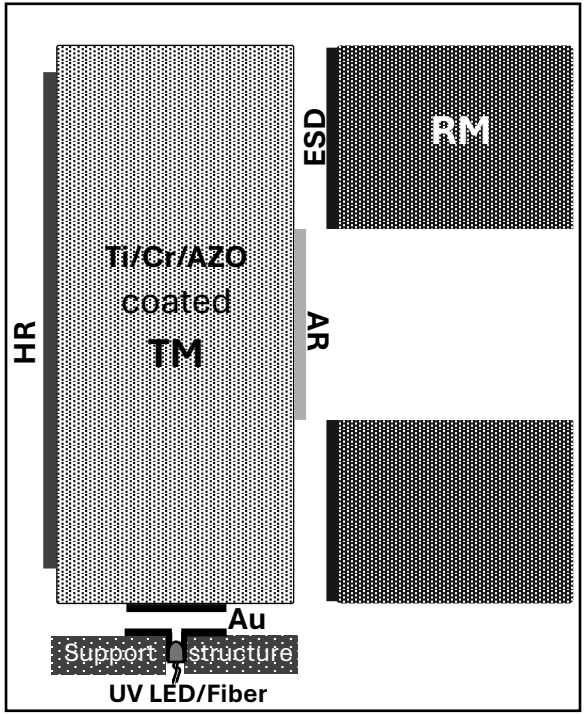

Figure 8. Schematic profile of the TM and RM with the UV discharge system and the Ti/Cr/AZO, HR, AR, ESD, and Au photoemission coatings

Completely coating the TM with a conducting layer avoids the potential for the accumulation of high electrostatic charges on its dielectric surfaces[7,9]. A proposed material for such an application is Aluminum-doped Zinc Oxide (AZO), an electrically conducting transparent film[37], that would cover the entire TM surface, including as an underlayer for the HR and the AR coatings. As development of AZO in conjunction with the HR and the AR coatings has not yet reached implementation level[38,39], we suggest the use of a conduction film of Ti, Cr, or AZO that would not cover the HR and AR surfaces. Figure 8 is a schematic profile of the TM and RM with the UV discharge system and their Ti/Cr/AZO, HR, AR, ESD, and photoemission Au coatings. The design of the ES to TM contact is the same as previously shown in Figure 6.

We approximate $C_{TM}^T$, the total capacitance to ground of the coated TM, as the sum of the $\leq 100$ pF parallel plate capacitance to the ESD ($\leq 500$ cm$^2$ area, 0.5 cm gap) and the $< 10$ pF TM capacitance to space, resulting in $C_{TM}^T \cong 100\text{pF}$. After discharging to $|V_{TM}| \leq 10\text{mV}$ the total TM charge is $|Q_{TM}| \leq 1\text{pC}$, meeting the goal of 1 pC.

Columns 8 to 10 of Table 3 give the discharge time constants for this configuration. Discharge durations from $Q_{TM} = \pm 1\text{nC}$ are about 30 s and 2 s respectively. Note that the linear and exponential contributions to the discharge time from $Q_{TM} = -1\text{nC}$ are approximately equal due to the relatively large $C_{TM}^T$ capacitance. As calculated from equation 2, $\Delta T < 20\text{mK}$ is the maximum temperature rise of the TM where we assumed $\epsilon_{TM} \cong 1$ for a conducting surface.

For a TM conducting surface opposite the RM, the ESD can be designed as a parallel plate capacitor (PPC), rather than the present LIGO interlaced fingers layout that generates the gradients required for the drive of a dielectric; see Figure 9a. The maximum force of the spiral ESD[40], at an activation voltage of $V_E^S = 800\text{V}$ and a gap $d_E = 5\text{mm}$ is $F_E^S = 450\text{μN}$, that we compare to the equivalent force per $V^2$ for a PPC and calculate the ratio of the voltages:

$$\frac{F_E^S}{V^2} = 0.7 \text{ nN/V}^2, \quad \frac{F_E^C}{V^2} = \frac{1}{2}\frac{\epsilon_0 A_E^C}{d_E^2} \cong 8.9 \text{ nN/V}^2, \quad F_E^C = F_E^S \Rightarrow \frac{V_E^C}{V_E^S} \cong \sqrt{\frac{0.7}{8.9}} = 0.28 \qquad 10$$

where the superscripts $S$ and $C$ refer to the spiral and PPC ESD configurations respectively and $A_E^C \cong 500\text{cm}^2$. Results from a finite element analysis, that aimed to optimize the spiral drive, validates the ratio of the voltages required to achieve the same force with a PPC[41] as $\leq 28\%$, or a PPC bias (preload) voltage $V_E^{CB} \cong 200V$. Figure 9b shows a conceptual design for a PPC ESD.

In order to null the charges induced on the coated configuration of the TM shown in Figure 8 by the PPC drive we require, using four equal capacitances $C_{Ei}^C = C = 25\text{pF}$, arranged as in Figure 9b:

$$\sum_{i=1}^{4} V_{Ei}^C C_{Ei}^C = 0 \implies \sum_{i=1}^{4} V_{Ei}^C = 0 \qquad 11$$

In Appendix A we give an example of the dependence of the bias and control voltages on the required TM horizontal displacement and tilt and azimuthal angles for an ESD in the PPC configuration shown in Figure 9b

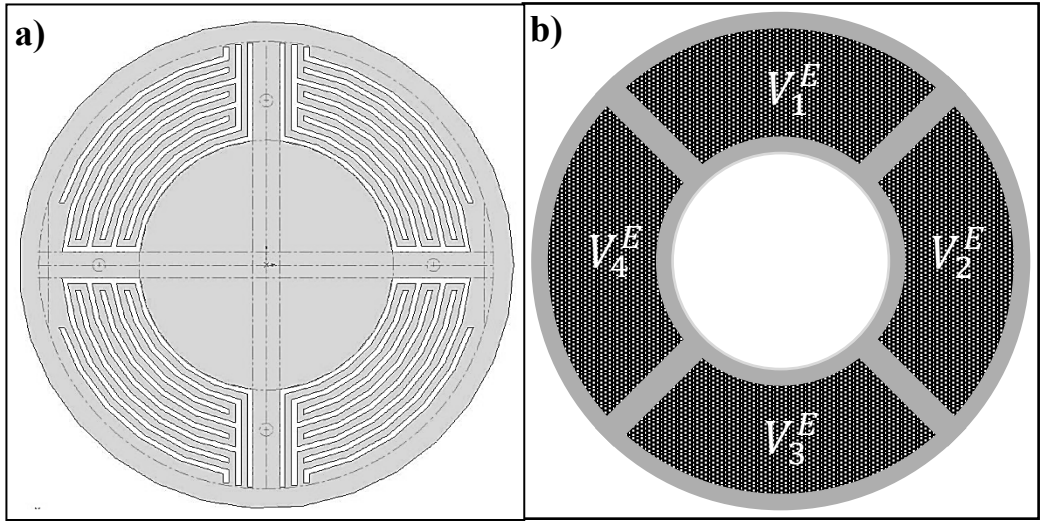

**Figure 9. a) LIGO type ESD drive, b) parallel plate capacitor type drive for conducting TM**

Illuminated by 1.0 mW optical power at 275 nm, the 'passive' discharge of two facing Au coated surfaces (with a small gap to radius ratio) follows a linear dependence for $|V_{TM}| \geq \mathbb{V}_{PE} \cong 0.2V$ and an exponential decrease thereafter for $|V_{TM}| \leq \mathbb{V}_{PE} \cong 0.2V$. The maximum photoelectric currents, generated by the direct and reflected beams, (superscripts $D$ and $R$), are $I_{TM}^D \cong 10\text{pA}$ and $I_{TM}^R \cong 0.5\text{pA}$, a 20:1 ratio that translates respectively into the charge reduction rates:

$$\frac{dQ_{TM}^D}{dt}(\text{pC/s}) = -10(\text{pC/s/mW}) \times P_{UV}(\text{mW}), \quad \frac{dQ_{TM}^R}{dt}(\text{pC/s}) = 0.5(\text{pC/s/mW}) \times P_{UV}(\text{mW}) \qquad 12$$

independent of the capacitance and directly proportional to the power $P_{UV}$. The corresponding time derivatives of the voltage are inversely proportional to the capacitance $C_{TM}$ to be discharged:

$$\frac{dV_{TM}^D}{dt}(\text{V/s}) = -10\left(\frac{\text{V} \cdot \text{pF}}{\text{s} \cdot \text{mW}}\right)\frac{P_{UV}(\text{mW})}{C_{TM}(\text{pF})}, \quad \frac{dV_{TM}^R}{dt}(\text{V/s}) = 0.5\left(\frac{\text{V} \cdot \text{pF}}{\text{s} \cdot \text{mW}}\right)\frac{P_{UV}(\text{mW})}{C_{TM}(\text{pF})} \qquad 13$$

In the exponential decrease regime, $|V_{TM}| \leq \mathbb{V}_{PE} \cong 0.2V$, the time constants $\tau^D$ and $\tau^R$ are:

$$\tau^D(s) = 0.2 \left(\frac{\text{s} \cdot \text{mW}}{\text{pF}}\right) \frac{C_{TM}(\text{pF})}{P_{UV}(\text{mW})}, \quad V_{TM}^D(t) = V_{TM}^0 e^{-t/\tau^D} \qquad 14a$$

$$\tau^R(s) = 1.5 \left(\frac{\text{s} \cdot \text{mW}}{\text{pF}}\right) \frac{C_{TM}(\text{pF})}{P_{UV}(\text{mW})}, \quad V_{TM}^R(t) = V_{TM}^0 e^{-t/\tau^R} \qquad 14b$$

The discharge durations $T^{D/R}$ - direct and reflected UV - from $\mp V_{TM}^0$ to $|V_{TM}^f| \leq 10\text{mV}$ are then:

$$T^D(s) \cong \frac{-V_{TM}^0}{dV_{TM}^D/dt} + 3\tau^D = \left[\frac{V_{TM}^0(\text{V})}{10\left(\frac{\text{V} \cdot \text{pF}}{\text{s} \cdot \text{mW}}\right)} + 0.6\left(\frac{\text{s} \cdot \text{mW}}{\text{pF}}\right)\right] \frac{C_{TM}(\text{pF})}{P_{UV}(\text{mW})} \qquad 15a$$

$$T^R(s) \cong \frac{+V_{TM}^0}{dV_{TM}^R/dt} + 3\tau^R = \left[\frac{V_{TM}^0(\text{V})}{0.5\left(\frac{\text{V} \cdot \text{pF}}{\text{s} \cdot \text{mW}}\right)} + 4.5\left(\frac{\text{s} \cdot \text{mW}}{\text{pF}}\right)\right] \frac{C_{TM}(\text{pF})}{P_{UV}(\text{mW})} \qquad 15b$$

directly proportional to the capacitance to be discharged and inverse proportional to the UV power.

Consequently, 0.2 mW of optical power at 275 nm, applied to the Au coated photoelectrons generating location, will reduce a charging potential of 1 kV to $\leq |10\text{mV}|$ in a duration of between 5 min and 75 min. Using Au coatings and the 275 nm UV wavelength, we implement the passive discharge method with low energy "slow" photoelectrons, that does not require bias potentials and that was extensively demonstrated experimentally[27,32].

## IV. CONCLUSIONS

We have discussed two implementations for the TM discharge: separate discharge actuators at each of the eight ESs and a single common UV discharge. The ES discharge method has the advantage of requiring a minimum of new TM coatings - tabs at the ES-to-TM potential contact locations – and the disadvantage of eight discharge systems. A common UV discharge requires only one system but does, however, require more extensive additional TM coatings: either minimal thin traces connecting the tabs or a maximal coating of all dielectric surfaces of the TM including, if practical, undercoats of the HR and AR coatings. A maximal TM conductive coating reduces or eliminates the possibility of high electrostatic potentials accumulating on the bare dielectric surfaces but requires a redesign of the ESD system. We have shown that a schematically new ESD with four electrodes enhances the activation force by a factor of about four, for the same applied voltage, and can control simultaneously the axial displacement and the tilt and azimuthal orientations of the TM with zero potential induced on it.

UV power to the discharge locations can be delivered by either local UV-LEDs or by optical fiber cables. Using UV-LEDs avoids power losses due to attenuation in the fibers and various couplings and requires only electrical connection into the vacuum. The main disadvantage is the positioning of active electrical devices and their power supply lines in close proximity to the TM, a problem solved if fiber optic cables are used and the active UV sources are located outside the detector vacuum chamber. Power losses in the fibers and the coupling to the UV sources are readily compensated by the very high powers UV-LEDs can supply at 275 nm[34].

At a delivered power of 0.2 mW, estimated discharge durations to $|V_{TM}| \leq 10\text{mV}$, are 5 min and 75 min for direct and reflected UV illumination respectively. A potential of $\leq 10$ mV corresponds to a TM charge of ≤ 1 pC, meeting the 1 pC goal and more than one order of magnitude less than the 10 pF allocated in the error budget for Advanced LIGO.

We propose the passive "slow" photoelectrons charge management approach for consideration as an addition, or possibly a replacement, of the present LIGO ion sprayer system. The exact choice of system implementation type will depend on the interfaces with the existing hardware and a detailed analysis of any new noise sources added to the error budget. Our initial study has not identified any such error sources.

**APPENDIX A: Required Control Voltages for a Parallel Plate Capacitance ESD Drive**

As a simple example we consider the control of $\delta, \theta, \varphi$, the horizontal displacement and the tilt and azimuthal angles, using the AC and/or DC voltages $V_i^\delta, V_i^\theta, V_i^\varphi$ added as adjustments to the bias voltage $|V_{Ei}^{CB}| \equiv V^B$. To satisfy equation 11, and after simplifying the notation by removing the subscripts/superscripts $E/C$, we set:

$$\left.\begin{array}{l} V_{1,3}^B = -V_{2,4}^B = V^B \approx 200\text{V} \\ V_{1,3}^\delta = -V_{2,4}^\delta = V^\delta \leq 20\text{V} \\ V_{1,3}^\theta = \pm V^\theta \leq |20\text{V}|, \ V_{2,4}^\theta = 0\text{V} \\ V_{2,4}^\varphi = \pm V^\varphi \leq |20\text{V}|, \ V_{1,3}^\varphi = 0\text{V} \end{array}\right\} \quad 16$$

with the 3 variables $V^\delta, V^\theta$, and $V^\varphi$ to be determined from the force equation for the displacement $\delta$ and the torque equations for the tilt and azimuthal angles $\theta$ and $\varphi$ ($\delta_B$ is the bias displacement and $\mathbb{C} \equiv C/2d_E$):

$$\left.\begin{array}{l} \mathbb{C} \sum_{i=1}^{4} \left(V_i^B + V_i^\delta + V_i^\theta + V_i^\varphi\right)^2 = \alpha_\delta(\delta_B + \delta) \\ \mathbb{C}\left[\left(V_1^B + V_1^\delta + V_1^\theta\right)^2 - \left(V_3^B + V_3^\delta + V_3^\theta\right)^2\right] = \alpha_\theta \mathcal{R} \theta \\ \mathbb{C}\left[\left(V_2^B + V_2^\delta + V_2^\varphi\right)^2 - \left(V_4^B + V_4^\delta + V_4^\varphi\right)^2\right] = \alpha_\varphi \mathcal{R} \varphi \end{array}\right\} \quad 17$$

where $\alpha_\delta, \alpha_\theta$, and $\alpha_\varphi$ are the ESD force/spring constants for the axial, the tilt, and the azimuthal displacements respectively (in the presence of the bias voltage $V^B$) and $\mathcal{R}$ is its radial torque arm. Equations 13 reduce to:

$$\left.\begin{array}{l} 4(V^B + V^\delta)^2 + 2\left[(V^\theta)^2 + (V^\varphi)^2\right] = \dfrac{\alpha_\delta}{\mathbb{C}}(\delta_B + \delta) \\ 4V^\theta(V^B + V^\delta) = \dfrac{\alpha_\theta \mathcal{R}}{\mathbb{C}} \theta \\ 4V^\varphi(V^B + V^\delta) = \dfrac{\alpha_\varphi \mathcal{R}}{\mathbb{C}} \varphi \end{array}\right\} \quad 18$$

By definition, $\alpha_\delta \delta_B = 4\mathbb{C}(V^B)^2$ and the solutions for $V^\delta, V^\theta$, and $V^\varphi$ are:

$$\left.\begin{aligned}
V^\delta &= \sqrt{\frac{1}{8}\left\{\frac{\alpha_\delta\delta}{\mathbb{C}} + 4(V^B)^2 + \sqrt{\left(\frac{\alpha_\delta\delta}{\mathbb{C}} + 4(V^B)^2\right)^2 - 32\left[\left(\frac{\alpha_\theta\theta\mathcal{R}}{4\mathbb{C}}\right)^2 + \left(\frac{\alpha_\varphi\varphi\mathcal{R}}{4\mathbb{C}}\right)^2\right]}\right\}} - V^B \\
V^\theta &= \alpha_\theta\theta\frac{\mathcal{R}}{4\mathbb{C}} \bigg/ \sqrt{\frac{1}{8}\left\{\frac{\alpha_\delta\delta}{\mathbb{C}} + 4(V^B)^2 + \sqrt{\left(\frac{\alpha_\delta\delta}{\mathbb{C}} + 4(V^B)^2\right)^2 - 32\left[\left(\frac{\alpha_\theta\theta\mathcal{R}}{4\mathbb{C}}\right)^2 + \left(\frac{\alpha_\varphi\varphi\mathcal{R}}{4\mathbb{C}}\right)^2\right]}\right\}} \\
V^\varphi &= \alpha_\varphi\varphi\frac{\mathcal{R}}{4\mathbb{C}} \bigg/ \sqrt{\frac{1}{8}\left\{\frac{\alpha_\delta\delta}{\mathbb{C}} + 4(V^B)^2 + \sqrt{\left(\frac{\alpha_\delta\delta}{\mathbb{C}} + 4(V^B)^2\right)^2 - 32\left[\left(\frac{\alpha_\theta\theta\mathcal{R}}{4\mathbb{C}}\right)^2 + \left(\frac{\alpha_\varphi\varphi\mathcal{R}}{4\mathbb{C}}\right)^2\right]}\right\}}
\end{aligned}\right\} \quad 19$$

For an estimate of the required activation voltages, accurate to better than 10%, we neglect the smaller second order terms, $(V^B)^2 \gg (V^\delta)^2 \approx (V^\theta)^2 \approx (V^\varphi)^2$, and obtain:

$$\left.\begin{aligned}
V^\delta &\cong \alpha_\delta\delta\frac{1}{8\mathbb{C}V^B} \\
V^\theta &\cong \alpha_\theta\frac{\mathcal{R}}{\mathbb{C}\left(V^B + \alpha_\delta\delta\frac{1}{8\mathbb{C}V^B}\right)}\theta \cong \alpha_\theta\theta\frac{\mathcal{R}}{2\mathbb{C}V^B} \\
V^\varphi &\cong \alpha_\varphi\varphi\frac{\mathcal{R}}{\mathbb{C}\left(V^B + \alpha_\delta\delta\frac{1}{8\mathbb{C}V^B}\right)} \cong \alpha_\varphi\varphi\frac{\mathcal{R}}{2\mathbb{C}V^B}
\end{aligned}\right\} \quad 20$$

In these simplified solutions of equations 19, the displacement and tilt and azimuthal angle corrections scale linearly with their control voltages, are independent of each other, and represent a good first order estimate of the required actuation voltages. The control output for the actual actuation voltages will correspond to the complete solutions of equations 19. For LIGO, Prokhorov and Kissel[42] give a detailed analysis of the one dimensional, axial only, ESD force and activation voltages, as well as discussing the TM charge measurement using the standard force modulation method[22], as well as the impact of charges on the support structure and the linearization of the force equations.

### ACKNOWLEDGEMENTS

The authors acknowledge the support by SN&N Electronics Inc. and thank Stanford University for the loan of laboratory equipment.

# REFERENCES

[1] B. P. Abbott, R. Abbott, T. D. Abbott, M. R. Abernathy, F. Acernese, K. Ackley, C. Adams, T. Adams, P. Addesso, *et al.* (LIGO Scientific Collaboration and Virgo Collaboration), *Observation of Gravitational Waves from a Binary Black Hole Merger*, Phys. Rev. Lett. **116** 061102 (2016)

[2] LIGO Laser Interferometer Gravitational-Wave Observatory

[3] VIRGO European Gravitational Observatory

[4] GEO600 Gravitational Wave Detector

[5] KAGRA Large-scale Cryogenic Gravitational Wave Telescope

[6] D. K. Davies, *Charge generation on dielectric surfaces*, J. Phys. D: Appl. Phys. **2(11)** 1533 (1969)

[7] R Weiss, *Note on Electrostatics in the LIGO suspensions*, LIGO document T960137-00-E (1995)

[8] V.B. Braginsky, O.G. Ryazhskaya, S.P. Vyatchanin, *Notes about noise in gravitational wave antennas created by cosmic rays*, Physics Letters A **350** 1 (2006)

[9] D. Ugolini, R. Amin, G. Harry, J. Hough, I. Martin, V. Mitrofanov, S. Reid, S. Rowan, K-X. Sun, *Charging Issues in LIGO*, Proceedings of the 30th International Cosmic Ray Conference, Mexico City, Mexico **3(2)** 1283 (2008)

[10] S. Rowan, S. Twyford, R. Hutchins and J. Hough, *Investigations into the effects of electrostatic charge on the Q factor of a prototype fused silica suspension for use in gravitational wave detectors*, Class. Quantum Grav. **14** 1537 (1997)

[11] L.G. Prokhorov and J. Kissel, *Interaction of the ESD with electrical charges of the test masses in Advanced LIGO*, LIGO Document G1600699 (2016)

[12] B. Lantz, *Estimate of force coupling between an optic with a uniform surface charge and the metal frame on the optic cage,* LIGO Document G1401179-v3 (2014)

[13] S. Reid, I. Martin, A. Cumming, W. Cunningham, J. Hough, P. Murray, S. Rowan, M. Fejer, A. Markosyan, R. Route, S. Hild, M. Hewitson, and H. Lück, *Charge Mitigation Studies*, LIGO Document G070605-00 (2007)

[14] V.P. Mitrofanov, L.G. Prokhorov, and K.V. Tokmakov, *Variation of electric charge on prototype of fused silica test mass of gravitational wave antenna*, Phys. Lett. A**300** 370 (2002).

[15] V. Mitrofanov, *Overview of Charging Research*, LIGO Document G080077-00-Z (2008).

[16] R. G. Maev and V. Leshchynsky, *Introduction to Low Pressure Gas Dynamic Spray: Physics & Technology 1st Edition*, Wiley-Vch Verlagsgesellschaft Mbh (2008).

[17] S. Buchman, R.L. Byer, D. Gill, NA Robertson, K-X. Sun, *Charge neutralization in vacuum for non-conducting and isolated objects using directed low-energy electron and ion beams*, Class. Quantum Grav. **25** 035004 (2008)

[18] R. Weiss, *Ionic Neutralization of Surface Charge on Mirrors*, LIGO Document G1000383-v1 (2010)

[19] R. Weiss, *Charging of the test masses past, present and future*, LIGO Document G1401153-v3 (2014)

[20] E. Sanchez, C. Torrie, R. Abbott, D. Coyne, *Test Mass Discharging System: Design Document*, LIGO Document T1400713-v2 (2014)

[21] M. Hewitson, H. Luck, H. Grote, S. Hild, S. Rowan, J. R. Smith, K. A. Strain, B. Willke, *Charge measurement and mitigation for the main test-masses of the GEO600 Gravitational Wave Observatory*, LIGO Document P070087-00-Z (2007)

[22] S. Buchman, T. Quinn, G.M. Keiser, D. Gill, and T.J. Sumner, *Charge measurement and control for the Gravity Probe B gyroscopes,* Rev. Sci. Instrum. **66(1)** 120 (1995)

[23] S. Buchman, J. A. Lipa, G.M. Keiser, B. Muhlfelder, J. P. Turneaure, *The Gravity Probe B gyroscope*, Class. Quantum Grav. **32(22)** 224004 (2015)

[24] S. Saraf *et. al.*, *Ground testing and flight demonstration of charge management of insulated test masses using UV LED electron photoemission*, Class. Quantum Grav. **33(2)** 245004 (2016)

[25] S. Buchman, T. S. M. Al Saud, A. Alfauwaz, R. L. Byer, P. Klupar, J. Lipa, C. Y. Lui, S Saraf, S. Wang, P. Worden, *Flight and ground demonstration of reproducibility and stability of photoelectric properties for passive charge management using LEDs*, Class. Quantum Grav. **40(2)** 025010 (2022)

[26] M. Armano, H. Audley, J. Baird, P. Binetruy, *et. al.*, (LPF Team) *Precision charge control for isolated free-falling test masses: LISA pathfinder results*, Phys. Rev. D **98**, 062001 (2018)

[27] S. Wang, S. Saraf, J. Lipa, D. Yadav, S. Buchman, *Two Approaches for the Passive Charge Management of Contactless Test Masses*, Class. Quantum Grav. **39** 195008 (2022)

[28] P. J. Wass, D. Hollington, T. J. Sumner, F. Yang, M. Pfeil, *Effective decrease of photoelectric emission threshold from gold plated surfaces*, Rev. Sci. Instrum. **90(6)** 064501 (2019)

[29] K-X. Sun, N. Leindecker, A. Markosyan, S. Buchman, R. Route, M. Fejer, R. Byer, H. Armandula, D. Ugolini, G. Harry, *Effects of Ultraviolet Irradiation on LIGO Mirror Coating*, LIGO Document G080150-00 (2008)

[30] F. Yang, Y. Bai, W. Hong, H Li, L. Liu, T. J. Sumner, Q. Yang, Y. Zhao, Z. Zhou, *Investigation of charge management using UV LED device with a torsion pendulum for TianQin*, Class. Quantum Grav. **37(11)** 115005 (2020)

[31] Seoul Viosys, UV CA3535 series (CUD7GF1A)

[32] H. Yu, M. H. Memon, D. Wang, Z. Ren, H. Zhang, C. Huang, M. Tian, H. Sun, and S. Long, *AlGaN-based deep ultraviolet micro-LED emitting at 275 nm*, Optics Letters **46(13)** 3271 (2021)

[33] Marktech Optoelectronics, TUD79H1B, TUD79F1B, TUD79B1B, TUD89F1B, TUD89H1B, TUD89B1B

[34] LedRise, Bolb S3535 UVC SMD LED 275nm 40mW

[35] TARLUZ Shanghai, UV Optimized Quartz Fiber

[36] Thorlabs, Ø200 μm Core Glass Clad Silica Multimode Optical Fiber, 0.22 NA

[37] I.R. Cisneros-Contreras, G. López-Ganem, O. Sánchez-Dena, Y.H. Wong, A.L. Pérez-Martínez, Rodríguez-Gómez A. *Al-Doped ZnO Thin Films with 80% Average Transmittance and 32 Ohms per Square Sheet Resistance: A Genuine Alternative to Commercial High-Performance Indium Tin Oxide*, Physics **5(1)** 45 (2023)

[38] A. Markosyan, *Fabrication, optical and electrical properties of slightly conductive Al-doped ZnO thin films: current and prospective studies*, LIGO Document G1500058-v1 (2015)

[39] A. Dana, A. Markosyan, R. Bassiri, E. Bonilla, B. Lanz, and M. Fejer, *Ultralow Absorption Conductive Coatings for Charge Dissipation in Vacuum Optics*, LIGO Document G2100552-v1

[40] K.A. Strain, *Electrostatic drive (ESD) results from GEO and application in Advanced LIGO*, LIGO Document T060015-00-K (2006)

[41] Z. Rose and G. Vajente, *Final Report: Finite Element Modeling of the electrostatic drive for C.Ri.Me. lab*, LIGO Document T1700275-v5 (2021)

[42] L.G. Prokhorov and J. Kissel, *Actuation force of the ESD with charges and cage effects*, LIGO Document T1500467 (2015)